\documentclass[twocolumn, showpacs,preprintnumbers,amsmath,amssymb]{revtex4}
\usepackage{amsmath,amssymb}
\usepackage{bm}
\usepackage{graphicx}
\usepackage{color}

\begin{document}

\title{Formation and collision of traveling bands \\
in interacting deformable self-propelled particles 
}
\author{Sadato Yamanaka$^{1,2}$ and Takao Ohta$^{1,3,4}$}
\affiliation{$^1$Department of Physics, Kyoto University, Kyoto, 606-8502, Japan \\
$^2$Institute of Industrial Science, The University of Tokyo, Tokyo 153-8505, Japan \\
$^3$Department of Physics, The University of Tokyo, Tokyo 113-0033, Japan  \\
$^ 4$Soft Matter Center, Ochanomizu University, Tokyo, 112-0012, Japan
}
\date{\today}

\begin{abstract}
We study  collective dynamics of interacting deformable self-propelled particles whose migration velocity increases when the local density of
particles is increased. Numerical simulations in two dimensions reveal that traveling bands similar to those found previously in the Vicsek-type model is easily formed by this  local density dependence of the migration velocity.  We show  that a pair of  stripe bands which are traveling to the opposite directions is not destructed by a head-on collision but survives again after collision.
\end{abstract}

\maketitle


\section{Introduction} \label{sec:Introduction}

Collective dynamics of self-propelled objects has been studied extensively for these almost two decades. Vicsek {\it et al.} introduced the seminal  dynamical model of point particles which tend to travel at constant velocity but to move in the same direction through the short range alignment  interaction \cite{Vicsek1995}. By adding random noises, numerical simulations were performed in two dimensions to investigate the transition from disordered to ordered states by increasing the density of particles and/or decreasing the noise intensity. 
The properties of transition have been investigated in detail for particles obeying the polar alignment interaction \cite{Gregoire04,Raynaud08,Chate2008,Peruani11,Farrell12}.
A coarse-grained continuum model has also been introduced for self-propulsion  \cite{Toner}. See the review articles  \cite{Vicsek2,Ramaswamy} and the earlier references therein. 


In the study of the  Vicsek-type model in two dimensions, Chat\'e and his  coworkers  \cite{Chate2008,Ginelli} and Bertin {\it et al.}  \cite{Bertin} found numerically a stable non-uniform state called a traveling band structure.  The uniform ordered state becomes unstable just before the transition to the disordered state.  Straight  bands of the high density ordered state appear in the disordered low density matrix in a system with periodic boundary conditions. All the particles in a band are traveling to the direction, on average,  normal to the boundary between the ordered and the disordered regions. Similar band structures have been found recently in a model of non-point rigid particles with a repulsive interaction \cite{Weber}.  It is also mentioned that traveling concentration waves have been observed experimentally in dense actin motility assays and by their numerical modeling  \cite{Schaller}.

Existence of a traveling band in the vicinity of transitions is one of the characteristic features in the order-disorder transitions far from equilibrium since no counterpart exist in thermal equilibrium. Traveling bands have also been investigated  in a hydrodynamic description of self-propelled particles \cite{Bertin,Peshkov}. These studies assume that the traveling velocity is constant. Recently, a similar band structure has been
  found in a hydrodynamic approach where the traveling velocity is an increasing function of the local density  \cite{Mishra1}. However, in the opposite case that  the migration velocity is decreased with increase of the local density, several inhomogeneous structures appear but no traveling bands are observed  \cite{Farrell12}. 

In the present paper, we investigate formation of traveling band structures in further details based on a particle dynamics. 
We consider the case that the migration velocity of individual particles is an increasing function of the local density. 
One of our concerns is the structural stability of the traveling band. Since it was found originally in the model system of point particles, we take into account  the excluded volume effect and deformability of particles and examine whether or not the band structure is stable against these degrees of freedom. 
Furthermore, we shall show that these bands are robust upon collision such that they are disturbed transiently during collision but survive again after collision recovering apparently the original shape.

In section II, we describe our model equations for the interacting deformable self-propelled particles  \cite{Itino2011,Itino2012}. The local-density dependence of the migration velocity is introduced. In section III, we carry out numerical simulations in two dimensions and show that traveling band structures are formed in the ordered state near the order-disorder transition. In section IV, numerical simulations of a pair of traveling bands in head-on collision are also carried out to confirm robustness of the bands just like solitons in integrable systems. Discussion is given in section IV. 

\section{Model equations} \label{sec:Model}

We consider an assembly of interacting deformable particles in two dimensions \cite{Itino2011,Itino2012}.  The time-evolution of the {\it i}-th particle obeys the following set of equations for the position of the center of mass $r^{(i)}_{\alpha}$, the velocity $v^{(i)}_{\alpha}$ and the deformation tensor $S^{(i)}_{\alpha\beta}$;
\begin{align}
     \frac{d}{dt} r_{\alpha}^{(i)} &= v_{\alpha}^{(i)} ,
     \label{eq:pos3}   \\
     \frac{d}{dt} v_{\alpha}^{(i)} &= \gamma(\rho^{(i)}) v_{\alpha}^{(i)}
                  - |{\bf v}^{(i)}|^2 v_{\alpha}^{(i)}
                  - a S_{\alpha \beta}^{(i)} v_{\beta}^{(i)}
                  + f_{\alpha}^{(i)}  ,
     \label{eq:vel3}   \\
     \frac{d}{dt} S_{\alpha \beta}^{(i)} &= - \kappa S_{\alpha \beta}^{(i)}
                  + b \left( v_{\alpha}^{(i)} v_{\beta}^{(i)}
                  - \frac{1}{2} |{\bf v}^{(i)}|^2
                  \delta_{\alpha \beta} \right)  ,
     \label{eq:tens3}
\end{align}
where $\kappa$, $a$ and $b$ are positive constants. The coefficient $\gamma$ depends on the local density as described below. The repeated greek indices imply summation. The tensor $S$ expresses elongation of a particle and is represented in terms of the unit normal  $\bm{n}$ parallel to the long axis of an elongated particle as
\begin{equation}
     S_{\alpha \beta}^{(i)} \equiv s_i \left( n_{\alpha}^{(i)} n_{\beta}^{(i)} - \frac{1}{2} \delta_{\alpha \beta} \right)  ,
     \label{eq:def_tens3}
\end{equation}
where $s_{i}(>0)$ is the magnitude of deformation.  When the  term $ f_{\alpha}^{(i)}$ 
is absent, the above set of equations is a model of a traveling particle which changes its shape depending on the
 migration velocity \cite{Ohta}.  The first two terms on the right hand side of eq.  (\ref{eq:vel3}) with positive constant $\gamma$ make the particle travel at a constant velocity. The term with the coefficient $a$ modifies the velocity when the particle is deformed. 
 The term with the coefficient $b$ in eq. (\ref{eq:tens3}) makes the particle elongate when the particle is traveling. When $b>0 (<0)$,  elongation is parallel (perpendicular) to the traveling velocity. There is a bifurcation by increasing the coefficient $\gamma$ or deceasing the values of $\kappa$ such that  the straight motion becomes unstable and a circular motion appears \cite{Ohta}.  In the present study shown below, we assume that $ b$ is positive and the individual particles undergo straight motion when the interaction between the particles is absent. 

Now, we consider the 
 interaction between a pair of two particles given by
the term ${ \bf f}^{(i)} $ in eq. (\ref{eq:vel3}), which  is the force acting on the $i$-th particle and takes the form 
\begin{equation}
{ \bf f}^{(i)} = K\sum\limits_{j=1}^N{\bf F}_{ij}Q_{ij} .
 \label{f}
\end{equation}
Here, the index $i$ is not summed over, 
$K$ is a positive constant, $N$ is the total number of particles, and 
\begin{equation}
{ \bf F}_{ij}= -\frac{\partial U({\bf r}_{ij})}{\partial{\bf r}_{ij}} ,
   \label{F}
\end{equation}
with ${\bf r}_{ij}={\bf r}^{(i)}-{\bf r}^{(j)}$.
The potential $ U({\bf r}_{ij})$ is assumed to be a Gaussian form as 
\begin{eqnarray}
U({\bf r}_{ij})=\exp\Big[-\frac{{\bf r}_{ij}^2}{2\sigma^2}\Big],
\label{U1}
\end{eqnarray}
with $\sigma=1$ throughout the present paper.
The other factor $Q_{ij}$ in eq. (\ref{f}) represents an alignment mechanism of the particles and is defined by
\begin{equation}
     Q_{ij} = 1 + \frac{Q}{2}\mathrm{tr} \left( S^{(i)}-S^{(j)} \right)^2  ,
     \label{eq:alignment1}
\end{equation}
where $Q$ is a positive constant. 
In two dimensions, we have
\begin{eqnarray}
 Q_{ij} 
&=&1+\frac{Q}{4}\Big[(s_i-s_j)^2 \nonumber \\
&+& 4s_is_j\sin^2(\theta^{(i)} - \theta^{(j)})\Big]  
\label{Q2}
\end{eqnarray}
from the parameterization ${\bf n}^{(i)}=(\cos\theta^{(i)}, \sin\theta^{(i)})$.
This indicates that when the elongated directions of two particles are parallel to each other, the repulsive interaction is weaker than that in a perpendicular configuration.  For simplicity, we do not consider random noises in eq.  (\ref{eq:vel3}) \cite{Itino2012}.

\begin{figure}[t]
      \begin{center}
       \includegraphics[width=0.2\textwidth]{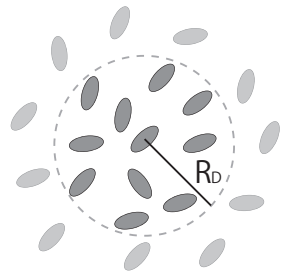}
      \end{center}
     \caption{Local density  defined by $N_i/\pi R_D^2$ for the $i$-th particle, where  $N_i$  is the number of the particles in the circular area with radius $R_D$ whose center is at the center of mass of the $i$-th particle. The $i$-th particle is not counted in $N_i$.}
     \label{fig:local_dens}
\end{figure}

\begin{figure}[t]
      \begin{center}
       \includegraphics[width=0.4\textwidth]{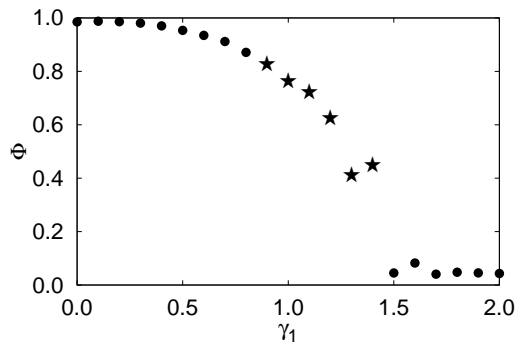}
      \end{center}
     \caption{Order-disorder transition by increasing the value of $\gamma_1$ for $\gamma_0=0.5$, $N=512$ and $\rho_0=0.06$. The traveling bands appear in the region indicated by the stars.}
     \label{fig:OP0}
\end{figure}

  \begin{figure}[t]
      \begin{center}
       \includegraphics[width=0.46\textwidth]{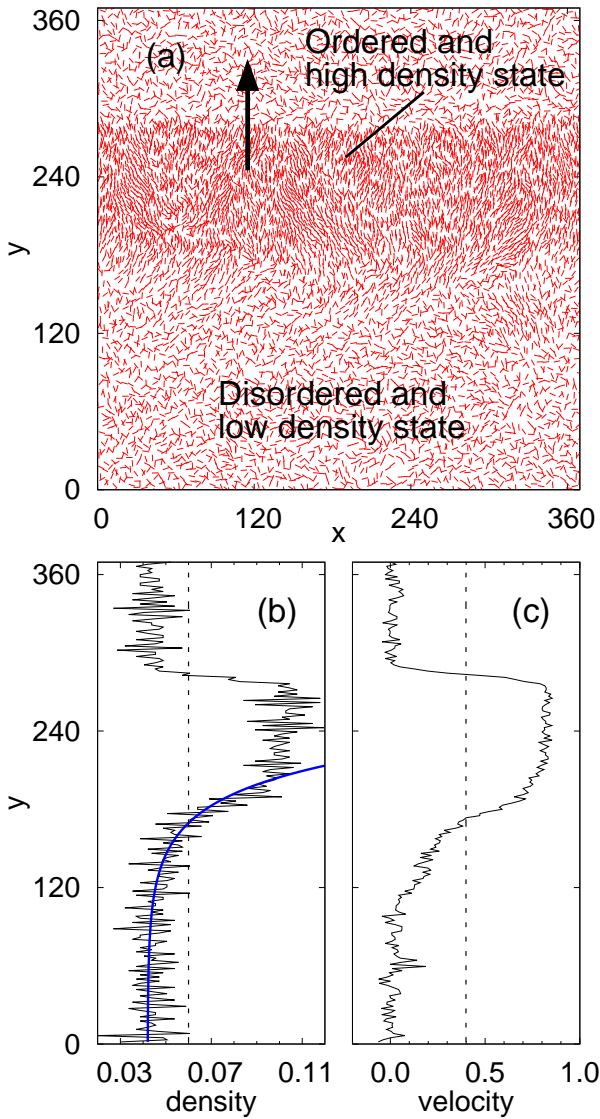}
      \end{center}
     \caption{(a) A band traveling upward  as indicated by the arrow for $\gamma_0=0.5$, $\gamma_1=1.3$, $N=8192$ and $\rho_0=0.06$. Each particle is displayed as a rod emphasizing the elongated direction. (b) Density profile along the  $y$  axis after averaged over the $x$ direction in (a).  The smooth full line indicates an exponential fitting at the rear of the band. The broken line indicates the average density $\rho_0$. (c)Velocity profile along the $y$ axis after averaged over the $x$ direction in (a). The broken line indicates the $y$-component of the traveling velocity average over the whole particles.}
     \label{fig:band}
\end{figure}

The coefficient $ \gamma(\rho^{(i)}) $ indicates that the migration velocity depends on the local particle density $\rho^{(i)}$. The definition of the local density around the $i$-th particle is given by
\begin{eqnarray}
\rho^{(i)}(t)=N_i(t)/(\pi R_D^2)  ,
 \label{eq:local}
\end{eqnarray}
where $N_i$ is the number of the particles within the distance $R_D$ from the center of mass of the $i$-th particle as depicted in Fig.   \ref{fig:local_dens}. 
The local density $\rho^{(i)}$ is time-dependent since the number $N_i$ is time-dependent due to collisions of particles. 
Here we consider the case that the velocity increases when the local density is increased
\begin{eqnarray}
 \gamma(\rho^{(i)}) =\gamma_0 + \gamma_1\left(\frac{\rho^{(i)}(t)}{\rho_0}-1\right)  ,
 \label{eq:defrho}
\end{eqnarray}
where $\gamma_0$ and $\gamma_1$ are positive constants and $\rho_0=N/(L_xL_y)$ with  $L_x$ and $L_y$ the 
linear dimensions of the rectangular system. It is worth mentioning that experiments in swimming bacteria have been reported recently where the average migration velocity increases with the density \cite{Sokolov}.

In order to characterize the ordered and the disordered states, we define the order parameter as
\begin{equation}
     \Phi = \left| \frac{1}{N} \sum_{j=1}^{N} e^{2i\theta^{(j)}} \right| .
     \label{eq:phi}
\end{equation}
This is the same as the nematic order parameter of elongated particles. We may define another order parameter by using the direction of migration velocity of each particle. However, we have checked that there is no appreciable difference in magnitude from that given by eq. (\ref{eq:phi}).

Before we describe our results, we briefly summarize the previous results obtained for the constant coefficient $\gamma$ without the local-density dependence \cite{Itino2011,Itino2012}. 
The ordered state appears by increasing the density. However, this ordered state becomes unstable for further increase of the density and a discontinuous transition to the disordered state occurs. That is, the ordered state appears in the finite region of the density. The ordered state also becomes unstable by increasing the noise intensity for fixed value of the density.  We have obtained a curious traveling band near the disorder-to-order transition threshold on the high density side by decreasing the density. This is an inverse band in a sense that  a high density disordered stripe is traveling in the matrix of the ordered state \cite{Menzel}. This emerges only for extremely large number of the particles as $N=$8192 and 32768 with $\rho_0=0.086$.
No stable inhomogeneous structures are observed in the order-to-disorder transition on the low density side.  Another kind of inverse traveling bands in which a disordered low density stripe is surrounded by the high density ordered state has been found in numerical simulations of polar disks \cite{Weber}.  

The above results imply that  formation of normal bands as Chat\'e {\it et al.} found \cite{Chate2008,Ginelli} numerically in the Vicsek-type model is unlikely to be observed in the set of eqs.  (\ref{eq:pos3}),   (\ref{eq:vel3}) and  (\ref{eq:tens3}) with the constant coefficient $\gamma$. However, the situation changes drastically if one allows a density dependence of $\gamma$ as we will show in the next section.


\section{Formation of traveling bands} \label{sec:Results1}
We have carried out numerical simulations of eqs. (\ref{eq:pos3}),   (\ref{eq:vel3}) and  (\ref{eq:tens3}) in a square area under periodic boundary conditions. The parameters are fixed as 
$\kappa=a=\sigma=1$, $b=0.5$, $K=R_D=5$ and $Q=50$. In most of simulations, we have put  $\rho_0=0.06$, and $N=512$ and $N=8192$ to check the system size effects. The results where the coefficient $\gamma_1$ is varied are shown in Fig.  \ref{fig:OP0} for $\gamma_0=0.5$ and $N=512$.  The value of the order parameter asymptotically in time is plotted as a function of $\gamma_1$. It is evident that the ordered state becomes unstable for $\gamma_1 \ge 1.5$. When $\gamma_1$ is large, small density fluctuations trigger formation of  clusters of propagating particles with high density, which is surrounded by motionless particles of low density. Since the motionless particles tend to be disk-shaped, the alignment mechanism (\ref{eq:alignment1}) contained in the interaction part does not work effectively and the system becomes a disordered state. 

In the ordered state, there are two different kinds of dynamics. 
When the coefficient $\gamma_1$ is smaller than  0.8, the homogeneous ordered state is evolved where all the particles are traveling, on average, to a certain direction.  However, in the  region $0.8<\gamma_1 <1.5$ indicated by the stars  in Fig.  \ref{fig:OP0}, this homogeneous state is unstable. A traveling band appears in which the particles are propagating normal to the stripe whereas the particles outside the bands are almost motionless and their positions are random.  In the small system of $N=512$, only one band appears but for larger systems, two or three bands  are often observed. A snapshot of the band structure is displayed in Fig.  \ref{fig:band}(a) where the band is traveling upward and the parameters are chosen as $\gamma_1=1.3$, $\gamma_0=0.5$, $N=8192$, $\rho_0=0.06$. The density profile and the velocity profile are also plotted. The change of the density and the velocity is steep in the front of the band and extended in the  rear. This tendency is the same as
  the one found by Chat\'e {\it et al.}  in the self-propelled point particles  \cite{Chate2008}. 

\begin{figure}[t]
      \begin{center}
             \includegraphics[width=0.4\textwidth]{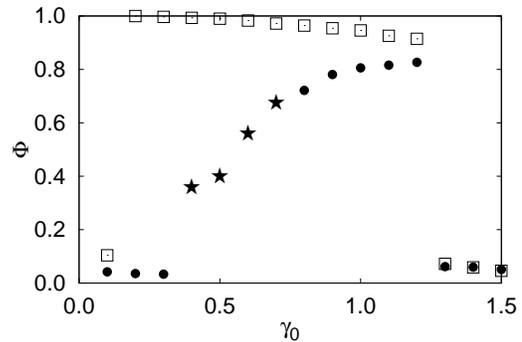}
      \end{center}
     \caption{Order parameter as a function of $\gamma_0$ for  $\gamma_1=1.3$ (dots). Traveling bands appear in the region indicated by the stars. For comparison, the case in the absence of the density dependence, i.e.,   $\gamma_1=0$ is also shown by the square symbols. The parameters are set to be $N=512$ and $\rho_0=0.06$. }
     \label{fig:OP1}
\end{figure}

 We have examined formation of bands by changing the  values of $\gamma_1$ as $\gamma_1 =0.9, 1.1$ and 1.3 for fixed values of $N=512$, $\rho_0=0.06$ and $\gamma_0=0.5$. Starting from random initial conditions, we have carried out 10 independent runs for each value of $\gamma_1$ and have found traveling bands for all simulations. 
 We have carried out alternative numerical simulations by changing the parameter $\gamma_0$ for  fixed values of the coefficient $\gamma_1$ as shown in Fig. \ref{fig:OP1}. The dots indicate the results for  $\gamma_1=1.3$. For comparison, the case without local-density dependence $\gamma_1=0$ is also plotted by the square symbols. In the absence of the local-density dependence of the migration velocity, the ordered state exists for $0.2 \le \gamma_0 \le 1.2 $. The ordered state does not appear for $\gamma_0<0.1$ since elongation of the particles is not large enough for small migration velocity so that orientational order of the particles cannot be achieved. When $\gamma_0$ is large, the straight motion becomes less stable, compared with a circular motion  \cite{Ohta} and therefore the alignment mechanism does not work effectively.  Another effect is that when $\gamma_0$ is large the kinetic energy is large compared with the interaction energy so that the alignment mechanism c
 ontained in the $Q$ term becomes comparably weak. This tendency can be seen also for $\gamma_1=1.3$. It is noted, however, that the degree of order is much smaller  in the region $0.4 \le \gamma_1 \le 0.7$ indicated by the stars where the traveling bands are observed. The properties of the bands are essentially the same as those in Fig. \ref{fig:band}.

In order to obtain more accurate results, numerical simulations have been carried out for $N=8192$. The data for the spatial profiles of the velocity and the local density are not scattered compared to those for $N=512$. We have found that  the maximum value of the density inside a band is not sensitive to the values of $\gamma_0$ and $\gamma_1$. Since the order parameter decreases  as the value of $\gamma_1$ is increased as shown in Fig.  \ref{fig:OP0}, this implies that the total area of bands decreases as the system approaches to the order-disorder transition threshold. 

Before closing this section, we make a remark about the front velocity of a band. We have estimated the front velocity from the numerical data and found that 
it increases as $\gamma_1$ is increased. For example, the  velocity is about $150\%$ of the average velocity of the particles inside the band for $\gamma_0=0.9$ whereas it is about  $180\%$ for $\gamma_0=1.3$ and for fixed values of $\gamma_0=0.5$ and $N=8192$. This is because the velocity of the particles in the vicinity of the front is more sensitive to the local density change when  $\gamma_1$ is larger. In fact, the front velocity does not depend on $\gamma_0$ appreciably for fixed values of $\gamma_1$. The front velocity of the bands in Fig. \ref{fig:OP1}  is larger than the average velocity of the inside particles by the factor of about $180\%$.

\begin{figure}[t]
      \begin{center}
       \includegraphics[width=0.4\textwidth]{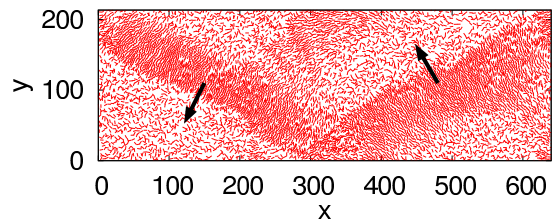}
      \end{center}
     \caption{Colliding bands traveling to the different directions for $\gamma_0=0.5$,  $\gamma_1=1.1 $  and $N=8192$. The arrows indicate the traveling directions.  These two bands travel persistently after collision. }
     \label{fig:complexband}
\end{figure}

\section{Collision of traveling bands} \label{sec:Results2}
In order to investigate the dynamics of bands carefully, numerical simulations for a rectangular system with the linear dimensions $L_x=640$ and $L_y=213.33$ are performed for $N=8192$ ($\rho_0=0.06$). Not only stripe bands parallel to the y-axis but also more complicated bands are often observed as shown in Fig. \ref{fig:complexband}.  These bands change their shape after collision but  are not destructed  and keep traveling  persistently.  

Collision of bands is investigated more systematically for a pair of bands traveling to the opposite directions. These bands are formed spontaneously starting with a random initial condition. In the initial transient regime, several bands, all of which are traveling to the same direction, appear. Occasionally (not always) the largest band splits into two due to large fluctuations in the tail part and the separated smaller one begins to travel to the opposite direction. Because of the periodic boundary conditions, these two bands get head-on collision and after each collision, the smaller band tends to grow in its size. Therefore these two bands become a comparable size eventually. The results shown below are the behavior obtained asymptotically in time. 

\begin{figure}[t]
      \begin{center}
       \includegraphics[width=0.4\textwidth]{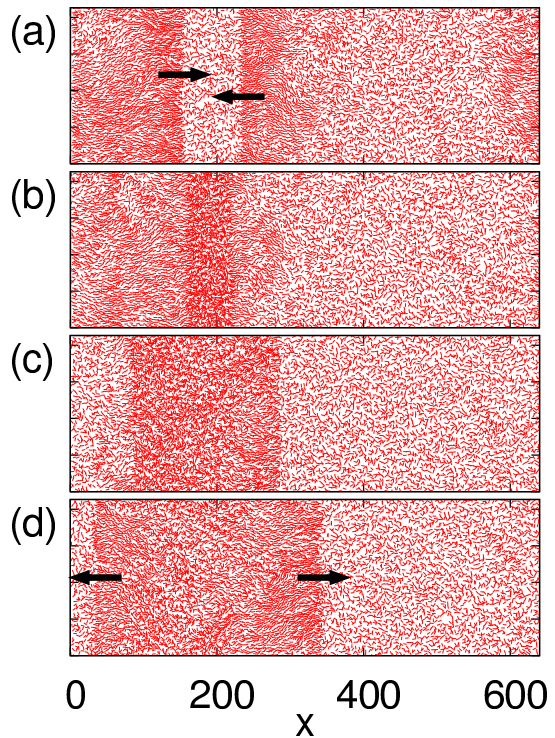}
      \end{center}
     \caption{Snapshots of head-on collision of a pair of bands for $\gamma_0=0.5$,  $\gamma_1=1.3 $  and $N=8192$
     at (a) $t=5636$, (b) $t=5676$, ( c) $5716$ and (d) $t=5756$.  The arrows indicate the traveling direction.}
     \label{fig:snapshot}
\end{figure}

\begin{figure}[t]
      \begin{center}
       \includegraphics[width=0.4\textwidth]{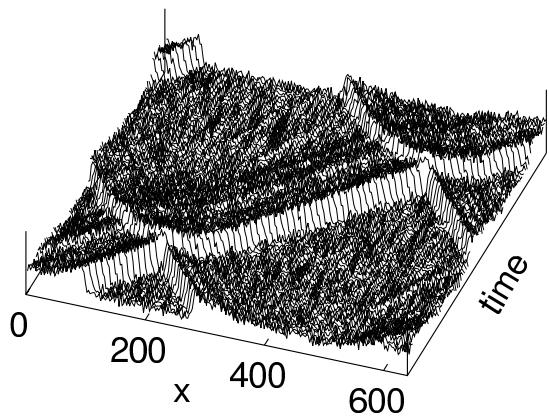}
      \end{center}
     \caption{Space-time plot of a pair of bands traveling to the opposite directions in the time interval from $t=5600$ to $t=6000$  for $\gamma_0=0.5$,  $\gamma_1=1.3 $  and $N=8192$.  }
     \label{fig:collision}
\end{figure}

Figure \ref{fig:snapshot} displays the snapshots just before and after collision. The orientational order in a band is destroyed transiently upon collision but there is a sharp front where  the density changes steeply as can be seen in Figs. \ref{fig:snapshot}(b) and (c ). As the front travels further, the orientational order is recovered near the front and the pair of bands is traveling away as in Fig. \ref{fig:snapshot}(d). 
Space-time plot of the local density averaged along the y-axis is shown in Fig. \ref{fig:collision}. It is evident that the bands survive again after collision and this phenomenon is repeated because of the periodic boundary conditions. 
We have checked that the two bands do not change their shape and speed appreciably even after seven consecutive collisions. 
One can see the tendency that the velocity of the bands is accelerated just after collision and then it becomes constant as the two bands are separated apart. 
We have observed preservation of traveling bands upon collision for $\gamma_1=$1.1 and 1.3 so far.

In ref. \cite{Chate2008}, 
Chat\'e {\it et al.} have shown reflection of a traveling band at the system boundary by imposing elastic collision between the point particles and the boundary. We emphasize that what we have found above has no relation with this because there are no rooms of elastic collisions in our system since the dynamics are dissipative with deformability of each particle. Solitary waves have been obtained theoretically in a hydrodynamic coarse-grained model \cite{Bertin, Gopinath}. However, stability upon collisions has not been investigated. Quite recently, Ihle has studied density instability in the Vicsek-type model in two dimensions by a kinetic approach   \cite{Ihle}. By solving numerically the kinetic equation of the probability function for the velocity direction and the particle position, it has been shown that  traveling bands are not destructed immediately by a head-on collision but only a bigger band survives  after  repeated encounters. Furthermore, the properties of the b
 and (solitary wave) seem to depend on the system size. Therefore, this observation is not related with the present  soliton-like behavior described above.

\section{Discussion} \label{sec:Discussion}
We have  investigated the order-disorder transitions of deformable self-propelled particles with repulsive interaction which contains an alignment mechanism. In particular, the migrating velocity of individual particles depends on the local density such that the velocity increases as the local density is increased. Traveling band structures appear in the ordered state in the vicinity of the order-disorder transitions. 
Both the density profile and the velocity profile are steep at the front and there is a plateau region followed by an exponential decay in the rear. 

The reason as to why band structures appear when the velocity is an increasing function of the local density can be understood qualitatively as follow. 
Let us suppose formation of a locally high density region due to fluctuations. Since the traveling velocity is higher in the region, the particles are further elongated and this enhances the orientational order in that region.  The particles catch up the slowly moving particles in front whereas the alingment effect is weaker for the particles in the rear where the density is lower since the front velocity is larger than that of each particle. When the high density region grows up to the system size, it constitutes a traveling band because of the periodic boundary conditions. 
Therefore, increase of the traveling velocity as the local density is increased is favorable for formation of stable traveling bands.  
If the local density dependence on the migration velocity is absent, formation of traveling bands is quite rare \cite{Itino2012}. Inverse bands are  possible only for extremely large system. This is because our model system has a repulsive interaction which tends to make the system uniform in density. Furthermore, deformability of individual particles prevents from alignment compared to the system of rigid rod particles. 

 As a remarkable property, we have found that the bands are quite robust upon head-on collision, which is very similar to solitons in integrable systems. 
We shall investigate, as a future problem,  the interaction of the traveling bands in further details by changing the parameters, for example, the softness of individual particles. 


\section*{Acknowledgements}  
We would like to thank Dr. Y. Itino for useful discussions for numerical simulations. 
This work was supported 
by Grant-in-Aids for Scientific Research A (No. 24244063) and  C (No. 23540449) from JSPS.


\begin{thebibliography}{15}

\bibitem{Vicsek1995}
T. Vicsek, A. Czir\'{o}k, E. Ben-Jacob, I. Cohen, and O. Shochet, Phys. Rev. Lett. {\bf 75}, 1226 (1995).

\bibitem{Gregoire04}
G. Gr\'egoire and H. Chat\'e, Phys. Rev. Lett. \textbf{92}, 025702 (2004). 

\bibitem{Raynaud08}
H. Chat\'e, F. Ginelli, G. Gr\'egoire, F. Peruani, and F. Raynaud, Eur. Phys. J. B \textbf{64}, 451 (2008). 

\bibitem{Chate2008}
H. Chat\'e, F. Ginelli, G. Gr\'egoire, and F. Raynaud, Phys. Rev. E \textbf{77}, 046113 (2008). 

\bibitem{Peruani11}
F. Peruani, F. Ginelli, M. B\"ar, and H. Chat\'e, J. Phys.: Conf. Ser. \textbf{297}, 012014 (2011). 

\bibitem{Farrell12}
F. D. C. Farrell, M. C. Marchetti, D. Marenduzzo, and J. Tailleur, Phys. Rev. Lett. \textbf{108}, 248101 (2012). 

\bibitem{Toner}
J. Toner and Y. Tu, Phys. Rev. Lett. {\bf 75}, 4326 (1995).

\bibitem{Vicsek2}
T. Vicsek and A. Zafeiris, Phys. Rep. {\bf 517}, 71 (2012).

\bibitem{Ramaswamy}
S. Ramaswamy, Annu. Rev. Condens. Matter Phys. {\bf 1}, 323 (2010).

\bibitem{Ginelli}
F. Ginelli, F. Peruani, M. B\"ar, and H. Chat\'{e}, Phys. Rev. Lett. {\bf 104}, 184502 (2010).

\bibitem{Bertin}
E. Bertin, M. Droz, and G. Gr\'{e}goire, J. Phys. A: Math. Theor. {\bf 42}, 445001 (2009).

\bibitem{Weber}
C. A. Weber, T. Hanke, J. Deseigne, S. L\'{e}onard, O. Dauchot, E. Frey and H. Chat\'{e}, Phys. Rev. Lett. {\bf 110}, 208001 (2013). 

\bibitem{Schaller}
V. Schaller, C.Weber, C. Semmrich, E. Frey, and A. R. Bausch,
Nature (London) {\bf 467}, 73 (2010).


\bibitem{Peshkov}
A. Peshkov, I. S. Aranson, E. Bertin, H. Chat\'{e}, and F. Ginelli, Phys. Rev. Lett. {\bf 109}, 268701 (2012).

\bibitem{Mishra1}
S. Mishra, A. Baskaran, and  M. C. Marchetti, Phys. Rev. E {\bf 81}, 061916 (2010).



\bibitem{Itino2011}
Y. Itino, T. Ohkuma, and T. Ohta, J. Phys. Soc. Jpn. {\bf 80}, 033001 (2011).

\bibitem{Itino2012}
Y. Itino and T. Ohta, J. Phys. Soc. Jpn. {\bf 81}, 104007 (2012).

\bibitem{Ohta}
T. Ohta and T. Ohkuma, Phys. Rev. Lett. {\bf 102}, 154101 (2009).

\bibitem{Sokolov}
A. Sokolov, I. S. Aranson, J. O. Kessler, and R. E. Goldstein, Phys. Rev. Lett. {\bf 98}, 158102 (2007).

\bibitem{Menzel}
M. Tarama, Y. Itino, A. M. Menzel, and T. Ohta, Euro. Phys. J. (submitted).

\bibitem{Gopinath}
A. Gopinath, M. F. Hagan, M. C. Marchetti and A. Baskaran, Phys. Rev. E {\bf 85}, 061903 (2012).

\bibitem{Ihle}
T. Ihle, arXiv:1304.0149v1 






\end{thebibliography}
\end{document}